\documentclass[11pt]{article} 
\usepackage{appb,epsfig,amssymb}

\newcommand{\bq}{\begin{equation}}
\newcommand{\eq}{\end{equation}}
\newcommand{\bqa}{\begin{eqnarray}}
\newcommand{\eqa}{\end{eqnarray}}
\newcommand{\ben}{\begin{enumerate}}
\newcommand{\een}{\end{enumerate}}
\newcommand{\bc}{\begin{center}}
\newcommand{\ec}{\end{center}}
\newcommand{\bqb}{\begin{eqnarray*}}
\newcommand{\eqb}{\end{eqnarray*}}

\def\pr#1#2#3{ Phys. Rev. ${\bf{#1}}$ (#2) #3}

\def\pl#1#2#3{ Phys. Lett. ${\bf{#1}}$ (#2) #3}

\def\np#1#2#3{ Nucl. Phys. ${\bf{#1}}$ (#2) #3}
\def\zp#1#2#3{ Z. f. Phys. ${\bf{#1}}$ (#2) #3}

\def\ie{{\it i.e.\/}}
\def\eg{{\it e.g.\/}}

\def\etal{{\it et.al.\/}}

\def\ol#1{\overline{#1}}

\def\L{ {\cal L }}
\def\O{ {\cal O }}

\def\R{ {\cal R }}

\def\mt{m_t}

\def\lam{\Lambda_{NP}}

\def\x{$\bar d^\gamma_1$}
\def\y{$\bar d^\gamma_2$}
\def\z{$\bar d^\gamma_3$}
\def\u{$\bar d^Z_1$}
\def\v{$\bar d^Z_2$}
\def\w{$\bar d^Z_3 $}

\begin{document}

\title{\large \bf SEARCH FOR  NEW PHYSICS THROUGH 
$e^- e^+ \to  t \bar t$
\thanks{Presented at The Cracow Epiphany 
Conference on Spin Effects in 
Particle Physics, Cracow, Poland, January 9-11, 1998.}
}
\author{G.J. Gounaris
\address{Department of Theoretical Physics, 
University of Thessaloniki,
\hspace{2.2cm}\null \\
Gr-54006, Thessaloniki, Greece.}
}
\maketitle 
\begin{abstract}
If all new particles are too heavy to be directly produced
in the future Colliders, then the long sought
New Physics (NP) could only appear in the form of  new
interactions, beyond those of the Standard Model (SM). 
Many of the processes that could be used to study these
interactions, have already been discussed in the
literature. Here, I first briefly discuss the list of all not
yet excluded CP conserving such 
interactions, realized  as $SU(3)\times SU(2) \times U(1)$ 
gauge invariant $dim=6$ operators affecting the Higgs
and the quarks of the third family.
 Subsequently, I concentrate on the $\gamma  t\bar t$ and 
$Zt\bar t$ vertices generated by NP,  and on the
possibility to study them by using  various spin  
asymmetries accessible in $e^-e^+ \to t \bar t$, for polarized
and unpolarized beams. It is found that these asymmetries can 
fully determine the form of the $\gamma t\bar t$ and $Zt\bar t$ 
couplings. 
\end{abstract}

  
\section{Introduction} 
The work I am presenting here was done in collaboration with
Fernand Renard, J. Layssac, D. Papadamou and M. Kuroda. 
It addresses the logical possibility that in the 
future Colliders there would
be no new particles discovered, apart from a standard-like 
Higgs boson. In such case, the only possibly observable NP
effect, would be induced by  
new interactions, beyond those expected in the SM. If the scale 
of NP is sufficiently high, the NP interactions could appear as
$SU(3)\times SU(2) \times U(1)$ gauge invariant
$dim=6$ operators, affecting the various particles present in 
SM \cite{Buchmuller}. \par

Since the Higgs is the 
most fascinating and mysterious of all fields appearing in SM, 
it seems quite plausible to assume that it is also the source
of NP. In such a case the above list of the NP induced operators
\cite{Buchmuller}, could be considerably reduced by assuming
that it  only involves  the Higgs, 
the quarks of the third family (for which SM suggests that they 
have the strongest affinity to Higgs), and of
course the gauge bosons inevitably introduced whenever
derivatives enter. The CP conserving such operators have been
listed in \cite{top-op, top-Young}, 
while the CP violating ones have appeared in \cite{top-CP, boson-CP}.
Below, we restrict ourselves only to the CP conserving NP 
operators.\par

In the present talk we   investigate the extent to
which polarization effects in  
$e^- e^+ \to t \bar t$ and the top 
decay amplitude $t \to b W$, can be used to study these NP 
operators. We only consider their 
dominant  contributions, which can 
only appear as modifications of the $\gamma tt$, $Ztt$ and $Wtb$
vertices. These arise either at tree level, or (in case this is
vanishing), at the 1-loop level; where it is sufficient to
retain only the leading-log part, whenever  it is enhanced by 
a positive power of $\mt$ \cite{GKR}. \par

\section{The New Physics Operators.}
We first turn to the purely bosonic CP conserving operators
\cite{Hagiwara-boson}. Discarding all operators that give
strongly constrained tree level contributions to LEP1
observables, we end up with the  operators 
\bqa
\O_W &= & {1\over3!}\left( \overrightarrow{W}^{\ \ \nu}_\mu\times
  \overrightarrow{W}^{\ \ \lambda}_\nu \right) \cdot
  \overrightarrow{W}^{\ \ \mu}_\lambda \ \ \
 ,  \ \  \label{listW} \\[0.1cm]  
\O_{W\Phi} & = & i\, (D_\mu \Phi)^\dagger \overrightarrow \tau
\cdot \overrightarrow W^{\mu \nu} (D_\nu \Phi) \ \ \  , \ \
\label{listWPhi} \\[0.1cm]
\O_{B\Phi} & = & i\, (D_\mu \Phi)^\dagger B^{\mu \nu} (D_\nu
\Phi)\ \ \  , \ \ \label{listBPhi} \\[0.1cm]
\O_{WW} & = &  (\Phi^\dagger \Phi )\,    
\overrightarrow W^{\mu\nu} \cdot \overrightarrow W_{\mu\nu} \ \
\  ,  \ \ \label{listWW}\\[0.1cm]
\O_{BB} & = &  (\Phi^\dagger \Phi ) B^{\mu\nu} \
B_{\mu\nu} \ \ \  , \ \ \   \label{listBB} \\[0.1cm] 
\O_G &= & {1\over3!}~  f_{ijk}~ G^{i\mu\nu}
  G^{j}_{\nu\lambda} G^{k\lambda}_{\ \ \ \mu} \ \ \
 ,  \ \ \   \label{listG} \\[0.1cm]
\overline{\O}_{DG} & =& 2 ~ (D_{\mu} \overrightarrow G^{\mu
\rho}) (D^{\nu} \overrightarrow G_{\nu \rho})  \ \ \
  , \ \  \label{listDG}  \\[0.1cm]  
\O_{GG} & = &  (\Phi^\dagger \Phi)\, 
\overrightarrow G^{\mu\nu} \cdot \overrightarrow G_{\mu\nu} \ \
\  ,  \ \    \label{listGG} \\[0.1cm] 
\O_{\Phi 2} & = & 4 ~ \partial_\mu (\Phi^\dagger \Phi)
\partial^\mu (\Phi^\dagger \Phi ) \ \ \  , \ \ \
  \label{listPhi2} \\[0.1cm] 
\O_{\Phi 3} & = & 8~ (\Phi^\dagger \Phi) ^3\ \ \  .
\   \label{listPhi3} 
\eqa
The effective lagrangian describing the NP contributions from
these operators may be written as
\bqa
{\L}_{bos} & = & \lambda_W {g\over M^2_W}\O_W+f_W{g\over2M^2_W}
\O_{W\Phi}+f_B{g'\over2M^2_W}  \O_{B\Phi}+ \nonumber \\
 & &\frac{d}{v^2} \ {\cal O}_{WW}+\frac{d_B}{v^2}
\ {\cal O}_{BB}+ \frac{f_{\Phi 2}}{v^2}
{\cal O}_{\Phi 2}~+~ \frac{d_G}{v^2} \ {\cal O}_{GG}\ \  \ \ .\ 
\label{listLbos}
\eqa
Unitarity considerations relate any non vanishing 
of these   NP couplings, to the corresponding $\lam$ scale
where it may be generated. This $\lam$
scale is \underline{uniquely} determined by  unitarity, and does
not depend on the normalization chosen to define the NP
coupling. Thus, using unitarity, we express the sensitivity
limits to the various NP scales defined in a physically appealing way. 
Referring to \cite{unit, HZGRV}, these relations are
\bq
|\lambda_W| \simeq 19~{M^2_W \over \lam^2} \ \ \ \ \ , \ \ \ \ \ \ \
|f_B| \simeq 98~{M^2_W\over\lam^2} \ \ \ \ \ , \ \ \ \ \
 \ |f_W| \simeq 31~{M^2_W \over\lam^2} \ \ \ \ ,\
\eq
\begin{eqnarray}
d & \simeq & \frac{104.5~\left ({\frac{M_W}{\Lambda_{NP}}}
\right )^2} {1+6.5 \left ({\frac{M_W}{\Lambda_{NP}}}\right )} \ \
\mbox{ for } d>0 \ , \nonumber \\
d & \simeq & -~ \frac{104.5~\left ({\frac{M_W}{\Lambda_{NP}}}
\right )^2} {1- 4 \left ({\frac{M_W}{\Lambda_{NP}}}\right )} \ \
\mbox{ for }\  d<0 \ , \  \\
d_B & \simeq & \frac{195.8 ~\left ({\frac{M_W}{\Lambda_{NP}}}
\right )^2} {1+200 \left  ({\frac{M_W}{\Lambda_{NP}}}
\right )^2}\ \
\mbox{ for } d_B>0 \ , \nonumber \\
d_B & \simeq & -~ \frac{195.8 ~\left ({\frac{M_W}{\Lambda_{NP}}}
\right )^2} {1 +50 \left  ({\frac{M_W}{\Lambda_{NP}}}
\right )^2}\ \
\mbox{ for }\  d_B<0 \ , \  
\eqa
while for $\O_{\Phi 2}$ we refer to \cite{HZGRV}, and
to \cite{ggHg} for $\O_{GG}$. \par

The operators in (\ref{listW}, \ref{listBPhi}) 
are best studied
through  $e^-e^+\to W^-W^+$ at LEP2 and linear
$e^-e^+$,  $\mu^-\mu^+$ colliders \cite{LEP2, LC200}.
Gauge and Higgs production in the $\gamma \gamma $ Collider, 
can give further information on $\O_W$, 
$\O_{W\Phi}$, $\O_{B\Phi}$, as well as on $\O_{WW}$ and
$\O_{BB}$. Important constraints on $\O_G$ and $\O_{DG}$
could arise by  dijet and multi jet production at the
upgraded Tevatron and LHC \cite{dijet}, while $\O_{GG}$ will need the
study the Higgs production at a hadron collider.\par

NP contributions from the bosonic operators to the 
$\gamma t t$, $Ztt$ and $Wtb$ vertices, studied through $e^-e^+
\to t \bar t$,  only appear 
1-loop level.  Thus, \eg\@ the dominant contribution from
$\O_{\Phi 2}$ gives  a purely axial $Ztt$ and a  
left-handed $Wtb$ coupling; while $\sigma_{\mu \nu}$ couplings
for $Ztt$ and $\gamma tt$  arise from $\O_{WW}$ and 
$\O_{BB}$ \cite{GKR}.\par

The NP operators containing  quarks of the
third family  are divided into three classes
\cite{GRVbb, top-op}. 
Class 1 contains the operators involving $t_R$,  
Class 2 those not involving $t_R$; while 
in Class 3 we put all operators involving covariant
derivatives of gauge boson field strengths and realated to the
currents through the equations of motion. The 
{\bf Class 1} operators, for the study of which the process 
$e^-e^+ \to t \bar t$ is most suited, are

\vspace*{0.5cm}
A1) \underline {Four-quark operators}
\bqa
\O_{qt} & = & (\bar q_L t_R)(\bar t_R q_L) \ \ \ , \ 
\label{listqt}\\[0.1cm] 
\O^{(8)}_{qt} & = & (\bar q_L \overrightarrow\lambda t_R)
(\bar t_R \overrightarrow\lambda q_L) \ \ \ ,\ \label{listqt8} \\[0.1cm]
\O_{tt} & = & {1\over2}\, (\bar t_R\gamma_{\mu} t_R)
(\bar t_R\gamma^{\mu} t_R) \ \ \ , \ \label{listtt}\\[0.1cm] 
\O_{tb} & = & (\bar t_R \gamma_{\mu} t_R)
(\bar b_R\gamma^{\mu} b_R) \ \ \ , \ \label{listtb}\\[0.1cm] 
\O^{(8)}_{tb} & = & (\bar t_R\gamma_{\mu}\overrightarrow\lambda t_R)
(\bar b_R\gamma^{\mu} \overrightarrow\lambda b_R) \ \ \ , 
\label{listtb8} \\[0.1cm]
\O_{qq} & = & (\bar t_R t_L)(\bar b_R b_L) +(\bar t_L t_R)(\bar
b_L b_R)\ \ \nonumber\\
\null & \null & - (\bar t_R b_L)(\bar b_R t_L) - (\bar b_L t_R)(\bar
t_L b_R) \ \ \ , \ \label{listqq}\\[0.1cm] 
\O^{(8)}_{qq} & = & (\bar t_R \overrightarrow\lambda t_L)
(\bar b_R\overrightarrow\lambda b_L)
+(\bar t_L \overrightarrow\lambda t_R)(\bar b_L
\overrightarrow\lambda  b_R)
\ \nonumber\\
\null & \null &
- (\bar t_R \overrightarrow\lambda b_L)
(\bar b_R \overrightarrow\lambda t_L)
- (\bar b_L \overrightarrow\lambda t_R)(\bar t_L
\overrightarrow\lambda   b_R)
\ \ \  . \label{listqq8} 
\eqa\\
B1) \underline {Two-quark operators.}
\bqa
\O_{t1} & = & (\Phi^{\dagger} \Phi)(\bar q_L t_R\widetilde\Phi +\bar t_R
\widetilde \Phi^{\dagger} q_L) \ \ \ ,\ \label{listt1} \\[0.1cm]
\O_{t2} & = & i\,\left [ \Phi^{\dagger} (D_{\mu} \Phi)- (D_{\mu}
\Phi^{\dagger})  \Phi \right ]
(\bar t_R\gamma^{\mu} t_R) \ \ \ ,\label{listt2} \\[0.1cm]
\O_{t3} & = & i\,( \widetilde \Phi^{\dagger} D_{\mu} \Phi)
(\bar t_R\gamma^{\mu} b_R)-i\, (D_{\mu} \Phi^{\dagger}  \widetilde\Phi)
(\bar b_R\gamma^{\mu} t_R) \ \ \ ,\label{listt3} \\[0.1cm]
\O_{D t} &= & (\bar q_L D_{\mu} t_R)D^{\mu} \widetilde \Phi +
D^{\mu}\widetilde \Phi^{\dagger}
(\ol{D_{\mu}t_R}~ q_L) \ \ \ , \label{listDt}\\[0.1cm] 
\O_{tW\Phi} & = & (\bar q_L \sigma^{\mu\nu}\overrightarrow \tau
t_R) \widetilde \Phi \cdot
\overrightarrow W_{\mu\nu} + \widetilde \Phi^{\dagger}
(\bar t_R \sigma^{\mu\nu}
\overrightarrow \tau q_L) \cdot \overrightarrow W_{\mu\nu}\ \ \
,\label{listtWPhi}\\[0.1cm] 
\O_{tB\Phi}& = &(\bar q_L \sigma^{\mu\nu} t_R)\widetilde \Phi
B_{\mu\nu} +\widetilde \Phi^{\dagger}(\bar t_R \sigma^{\mu\nu}
 q_L) B_{\mu\nu} \ \ \ ,\label{listtBPhi}\\[0.1cm]
\O_{tG\Phi} & = & \left [ (\bar q_L \sigma^{\mu\nu} \lambda^a t_R)
\widetilde \Phi
 + \widetilde \Phi^{\dagger}(\bar t_R \sigma^{\mu\nu}
\lambda^a q_L)\right ] G_{\mu\nu}^a  \ \ \ , \ \label{listtGPhi}
\eqa
while the corresponding NP effective lagrangian is
 written as
\bq
\L_{top} =  \sum_i { f_i \over m^2_t}\,\O_i \ \ \ . \ 
\label{listLtop}
\eq
None of the Class 1 operators is strongly constrained by presently
existing measurements. The unitarity constraints,
relating in a normalization-independent way, these NP couplings 
to the corresponding $\lam$ scales, are given in \cite{top-op}.\par

Among the 14 operators in (\ref{listqt}, \ref{listtGPhi}),
only $\O_{t2}$, $\O_{Dt}$, $\O_{tW\Phi}$ and $\O_{tB\Phi}$
give tree level contributions to the gauge vertices. 
1-loop $\gamma tt$ and $Ztt$ contributions of the 
$V$ and $A$ type are 
induced by $\O_{qt}$, $\O_{qt}^{(8)}$ $\O_{tt}$ and $\O_{tb}$;
while $\sigma_{\mu \nu}$ couplings
arise from  $\O_{tG\Phi}$, $\O_{qq}$ and $\O_{qq}^{(8)}$
\cite{GKR}. The sensitivity of $e^-e^+ \to t\bar t$ to these
operators is discussed in Table 1.\par

\section{Observables}
As already stated, within our approximations, the 
only relevant NP effects 
induced by the above operators, are expressed as 
contributions to the $\gamma tt$, $Ztt$ and $Wtb$
vertices. If CP is conserved, the $V t\bar t
~(V=\gamma,~Z)$ vertex is written as \cite{GLR-dj, treview,
tchiral, GKR}
\bqa  
\label{eq:dgz}
&& -i \epsilon_\mu^V J^{\mu}_V =  \nonumber \\
&& -i e_V \epsilon_\mu^V \bar
u_t(p)\left [\gamma^{\mu}d^V_1(q^2)+\gamma^{\mu}\gamma^5d^V_2(q^2)
+(p-p^{\prime})^{\mu}\frac{d^V_3(q^2)}{m_t}\right ] 
v_{\bar t}(p^{\prime}) , 
\eqa
where the normalization is 
determined by $e_{\gamma}\equiv e$ and $e_Z\equiv e/(2s_Wc_W)$,
while $d^V_i$ are in general $q^2$ dependent form factors. 
Similarly the $t(p_t) \to W^+(p_W) b(p_b) $ amplitude is written
as
\bqa
\label{eq:dw} 
&& -i \epsilon^{W*}_\mu J_W^{\mu}= 
 -i {g V^*_{tb}\over2\sqrt2}\epsilon^{W*}_\mu \bar
u_b(p_b) \cdot \nonumber \\
&&
[\gamma^{\mu}d^W_1+\gamma^{\mu}\gamma^5d^W_2+(p_t+p_b)^{\mu}d^W_3+
(p_t+p_b)^{\mu}\gamma^5d^W_4]u_t(p_t)\ .
\eqa\par

We consider the case where $t\bar t$ are produced in $e^-e^+\to
t \bar t$, and subsequently one of them decays semileptonicly;
say \eg\@\@ $t \to b W \to b l^+ \nu$, while $\bar t$ decays
purely hadronicly. The $e^-e^+ \to t \bar t$
amplitude at a scattering angle $\theta$, for $L$ or 
$R$ polarized $e^-$ beam, determines the density matrix 
$\rho^{L,R}_{\tau \tau^\prime}(\theta)$ of the
produced $t$ quark, which contains
all possible information on the production mechanism and NP 
expressed through 
the six couplings $(d^\gamma_j, d^Z_j)$.\par

 The subsequent
decay $t \to b W \to b l^+ \nu$ is determined by the 
decay  functions $\R_{\tau \tau^\prime}$ which contain all
NP dynamical information coming from the four $d^W_i$ couplings;
and depend also on the three Euler angles $(\varphi_1,
\vartheta_1, \psi_1)$ determining the top decay plane in its rest
frame, as well as on the angle $\theta_l$ describing  the
angular distribution of $l^+$ in the top decay plane.\par

We thus have  \cite{GKR}) 
\begin{equation}
\frac{d\sigma^{L, R}(e^-e^+ \to b l^+ \nu)}
{d\cos\theta d\varphi_1 d\cos\vartheta_1 d\psi_1
d\cos\theta_l} \sim 
\rho^{L, R}_{\tau \tau^\prime}(\theta) 
\cdot {\cal R }_{\tau \tau^\prime }( \varphi_1 , \vartheta_1,
\psi_1, \theta_l) ,
\label{eq:dsigma}
\end{equation}
where the upper index describes the longitudinal polarization
$e^-$ beam. Eqs. (\ref{eq:dsigma}) may be written as
\begin{eqnarray}
\label{eq:rhoR}
\rho^{L , R}_{\tau_1 \tau_2} \cdot {\cal R }_{\tau_1 \tau_2}&=
&\frac{1}{2}~
(\rho_{++}+\rho_{--})^{L , R}( {\cal R }_{++}+ {\cal R }_{--})
\nonumber \\
&+&\frac{1}{2} (\rho_{++}-\rho_{--})^{L, R}( {\cal R }_{++}- {\cal R }_{--})
\nonumber \\
& +&~ \rho^{L, R}_{+-}( {\cal R }_{+-}+ {\cal R }_{-+})\ \  ,
\end{eqnarray}
in which the three terms in the r.h.s of (\ref{eq:rhoR}),
called respectively $A$, $H$ and $T$ terms,  can be
separated  by averaging (\ref{eq:dsigma}) in the Euler-angle
space, using the weights \cite{GKR, GLR-dj}
\bqa
A \sim  (\rho_{++}+\rho_{--})^{L , R}& 
\longleftarrow & 1 \cdot d\varphi_1 d\cos\vartheta_1
d\psi_1 \ , \label{A1} \\
H \sim (\rho_{++} -\rho_{--})^{L , R}& 
\longleftarrow & (\cos \psi_1 +r \sin \psi_1) \cdot
d\varphi_1 d\cos\vartheta_1  d\psi_1 \ , \label{H1} \\
T\sim \rho^{L, R}_{+-} & \longleftarrow & 
\Bigg \{ (\cos\psi_1 \sin \varphi_1 \cos
\vartheta_1 -\sin\psi_1 \sin \varphi_1)+ \nonumber \\
&& r (\sin\psi_1 \cos\varphi_1 \cos\vartheta_1 + \cos\psi_1
\sin\varphi_1 ) \Bigg \} \cdot  \nonumber \\
&& d\varphi_1 d\cos\vartheta_1  d\psi_1 \ ,
\label{T1} 
\eqa
where we choose   
\bq
\label{eq:r}
 r\equiv {3\pi m_t M_W\over4(m^2_t-2M^2_W)} \ \ ,
\eq
so  that to maximize the statistical significance of the 
results. \par

Integrating further over $\cos \theta$, we can then construct
from (\ref{A1}-\ref{T1}) $\theta_l$-asymmetries supplying
information  
on the top decay NP. Unfortunately, to first order in the NP
couplings, these asymmetries only measure the combination 
$d^W_3+d^W_4$; while the $t$ width is to the same order only
sensitive to $(d^W_1-d^W_2)^{NP}$. Thus, as far as the NP
affecting the $t$ decay is concerned, this method can at most
provide two independent constraints on the four possible
$d^W_i$ couplings.\par
 
If instead we integrate (\ref{A1}-\ref{T1}) over $\theta_l$,
then the $\rho^{L\pm R}_{\tau tau^\prime}$ in the r.h.s.
of (\ref{A1}-\ref{T1}) are determined as functions od $\theta$,
so that    
\begin{eqnarray}
(\rho_{++}+\rho_{--})^{L\pm R} & = & 2 e^4\left [\sin^2\theta
 \left({\frac{4 m^2_t}{s}}\right ) A^{L\pm R}_1
+(1+\cos^2\theta)A^{L\pm R}_2 -  4\beta_t \cos\theta
  A^{L\pm R}_3 \right ] ,\ \ \ \label{eq:rhotrace} \\[0.1cm]
(\rho_{++}-\rho_{--})^{L\pm R}& = & 4 e^4[(1+\cos^2\theta) 
~\beta_t  B^{L\pm R}_1
-  \cos\theta B^{L\pm R}_2] \ , \ \label{eq:rho++m--} \\[0.1cm]
 \rho^{L\pm R}_{+-}& = & e^4\left ( \frac{4\mt}{\sqrt{s}} \right )
\sin\theta \left[  C^{L\pm R}_1 - \cos\theta ~\beta_t 
C^{L\pm R}_2 \right ]
 \ \ , \label{eq:rho+-}
\eqa
with  $\beta_t$ being the top velocity, and 
$A^{L\pm R}_j$, $B^{L\pm R}_j$, $C^{L\pm R}_j$ depending only on
the production dynamics. \par

To first order in NP, we have
$B^{L\pm R}_2 \simeq A^{L\mp R}_2$ and 
$B^{L\pm R}_1 \simeq A^{L\mp R}_3$, which through  
(\ref{eq:rhotrace}-\ref{eq:rho+-}) means that in the
case of \underline{polarized beams},
we can construct  11 different asymmetries, 
as well as
measuring the overall magnitude of the $\theta$ integrated 
three independent $\rho^{L+R}_{ij}$ matrix elements, \ie\@
14 independent observables are accessible.
In the case of \underline{unpolarized} beams we can measure
instead 4 asymmetries and in addition the three independent 
$\rho^{L+R}_{ij}$ matrix elements as above; \ie\@
7 independent observables.\par

\section{Results and Conclusions}

Defining the NP contribution to the $\gamma tt$ and $Ztt$
couplings by (\ref{eq:dgz}) and
\bq
\ol{d}^V_j ~=~ d^V_j -d^{V,\, SM}_j
\eq
we  construct the $1\, \sigma$ ellipsoid in the  6-parameter
$\ol{d}^\gamma_j, ~\ol{d}^Z_i$ space, assuming only statistical
uncertainties for the various angular asymmetries, but reducing 
the overall number of event by a factor of $18\%$ due to branching
ratios, reconstruction of events and efficiencies \cite{effic, GLR-dj}. 
In addition
for the measurement of the integrated 
$\rho^{L+R}_{++}$, $\rho^{L+R}_{--}$ and $\rho^{L+R}_{+-}$
matrix elements we consider two cases with  additional
uncertainties; (a) $\sim 2\%$, (b) $20\%$ \cite{GLR-dj}. 
The $e^-e^+$ luminosity is taken 
$\L=80 fb^{-1}(s/TeV^2)$. In Figs.1-3 we present the results for the
projection of the 6-parameter ellipsoid to various 2-parameter
subspaces. The left half of the figures is always for
polarized beams, while the right for unpolarized
ones.\par 

Finally in Table 1, we give the sensitivity limits in the case
that only one of the above defined operators contributes at a
time, for a $0.5,~1$ or $2TeV$ Collider. In each case,
the sensitivity limits are
translated to lower bounds on the corresponding $\lam$ 
scales, using unitarity.

The conclusions reached through the above analysis are:
\begin{itemize}
\item
The top-spin characteristics of the $e^-e^+ \to
t \bar t \to (bl^+\nu) \bar t $ production process analyses
powerfully the NP contributions to the $\gamma tt $ and $Z tt$
vertices. A sufficient number of observables exist to constrain
all NP couplings. 
\item
Contrary to what happens in the $W^+W^-$ production
case, increasing the Collider energy does not always increase
the sensitivity to NP. Among  the couplings defined
in (\ref{eq:dgz}), it is mainly for the $d^\gamma_3$ and $d^Z_3$
couplings, that the sensitivity increases with energy. 
\item
For the $t \to b W$ decay, sensitivities are appreciable only for
the combinations $(d^W_3+d^W_4)^{NP}$ and  $(d^W_1-d^W_2)^{NP}$.
\item
From the fact the magnitudes of the left and right ellipses in
many cases in Figs.1-3 are rather similar, we conclude that 
$e^\pm$ polarization is  generaly \underline{not} very
important. 
\item
Dynamical models suggest that processes like
$e^-e^+ \to t\bar t$, as well as $e^-e^+ \to HZ, ~H\gamma$ or
$\gamma \gamma \to HZ$, are more promisings than $e^- e^+ \to
W^-W^+$, for detecting New Physics, in the case that no new
particles are produced in the future Colliders \cite{top-op}. 
\end{itemize}


\begin{figure}[p]
\vspace*{-2.cm}
\[
\epsfig{file=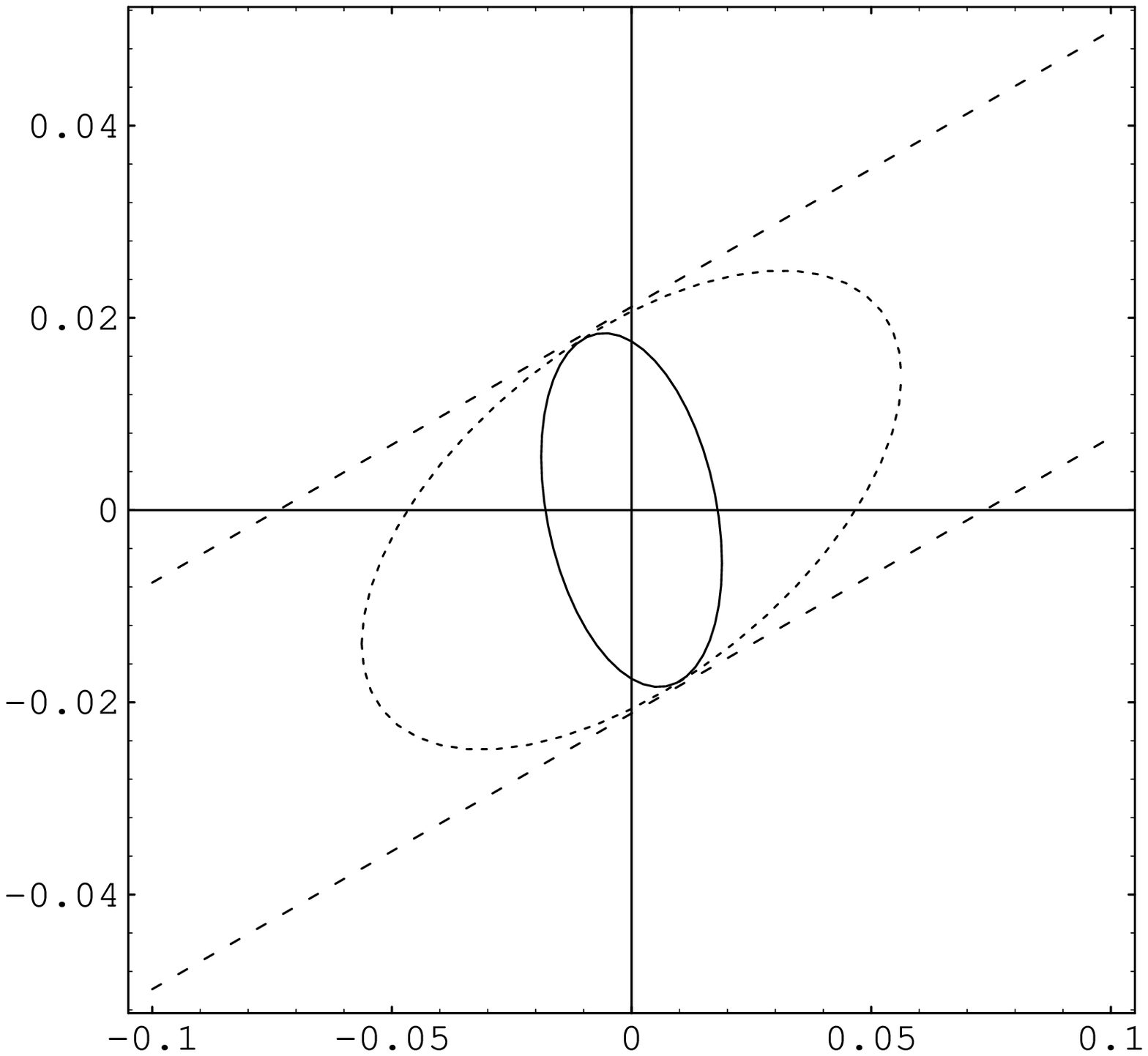,height=9cm}\hspace{2cm} 
\epsfig{file=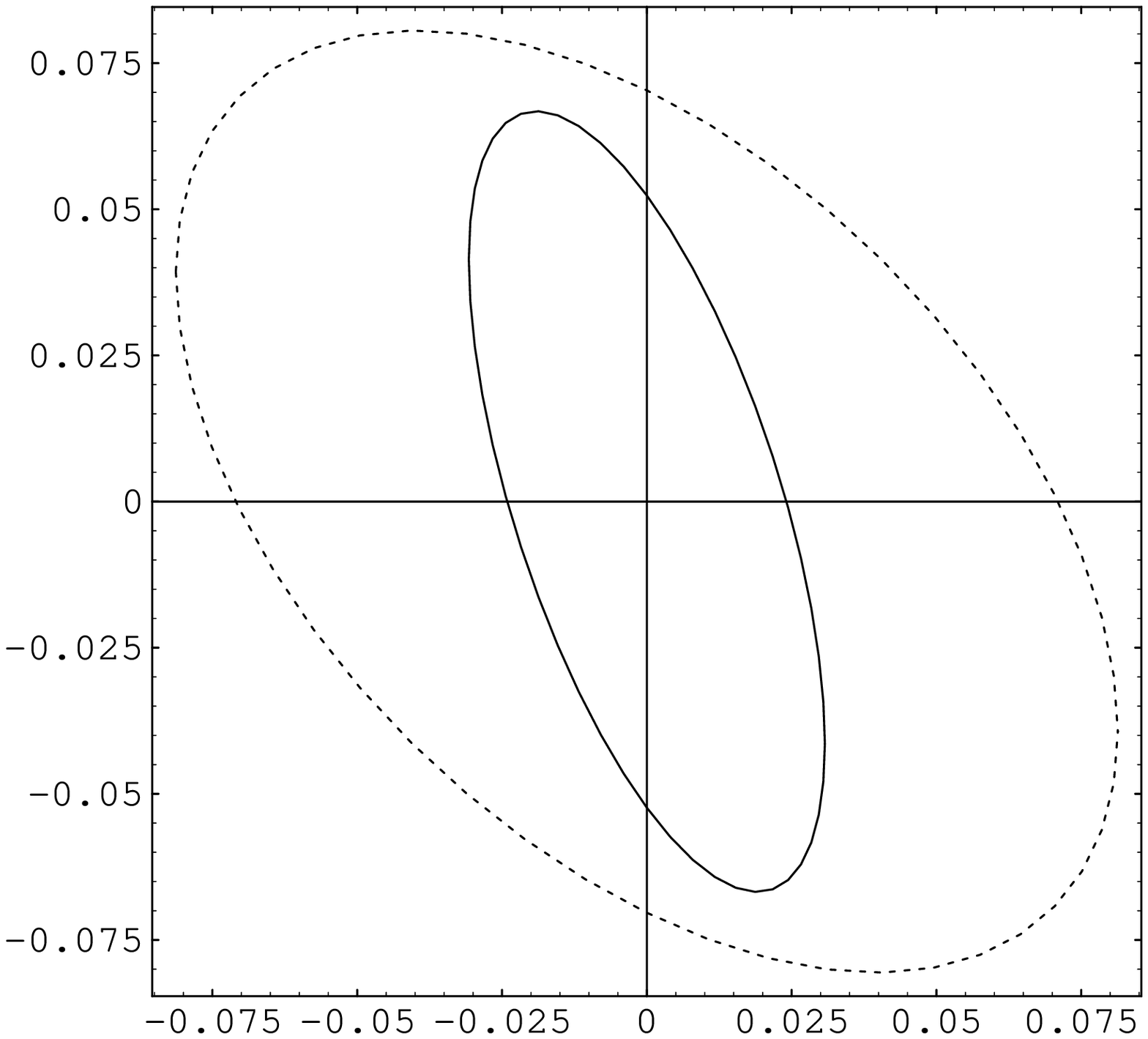,height=9cm}
\]
\vspace{-6.cm}\null\\
\hspace*{-1cm} \u \hspace{8cm} \u \\[3.2cm]
\hspace*{6cm} \x \hspace{8cm}  \x
\\
\hspace*{3.2cm} (a) \hspace{7.8cm}  (b)
\[
\epsfig{file=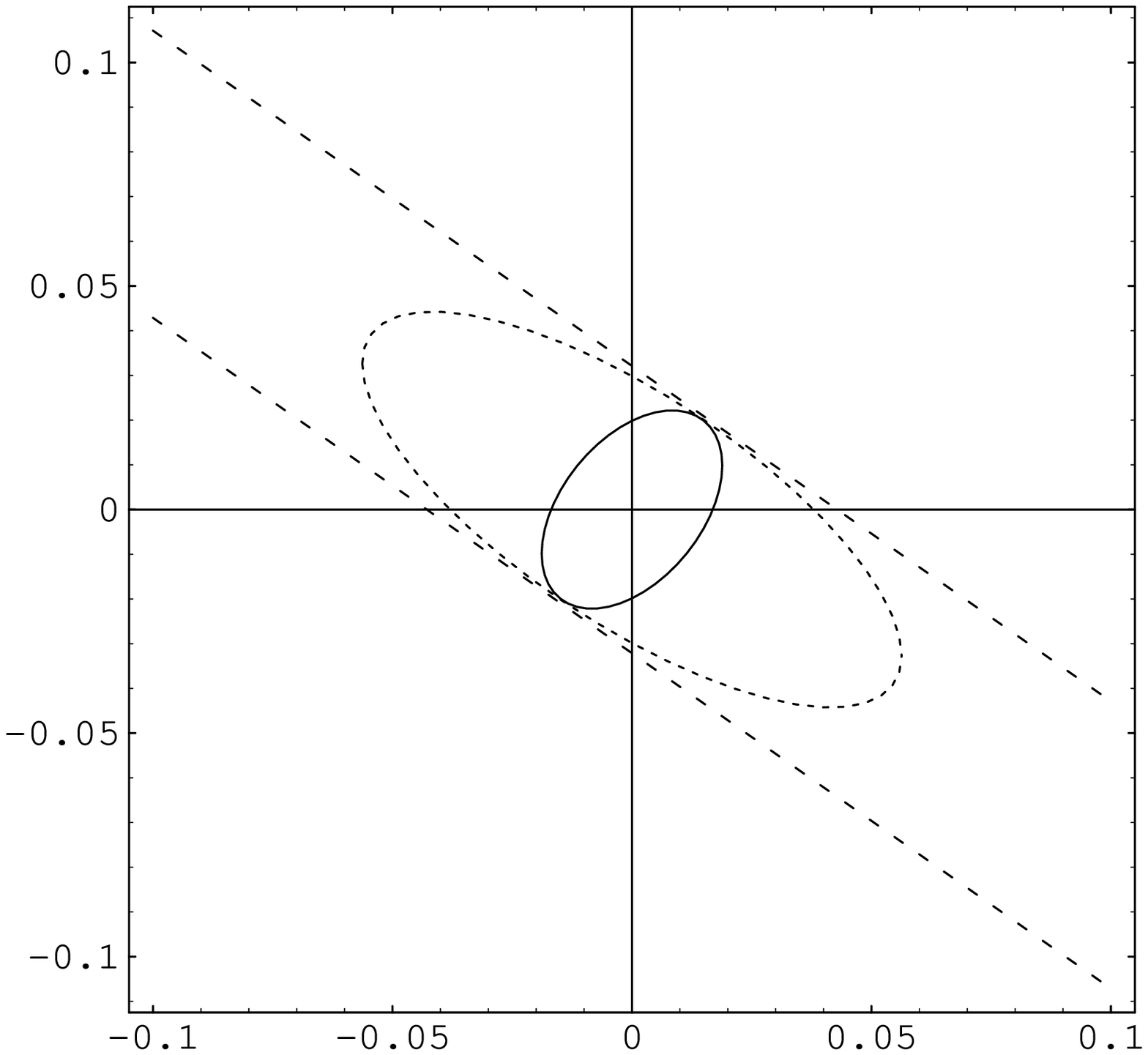,height=9cm}\hspace{2cm} 
\epsfig{file=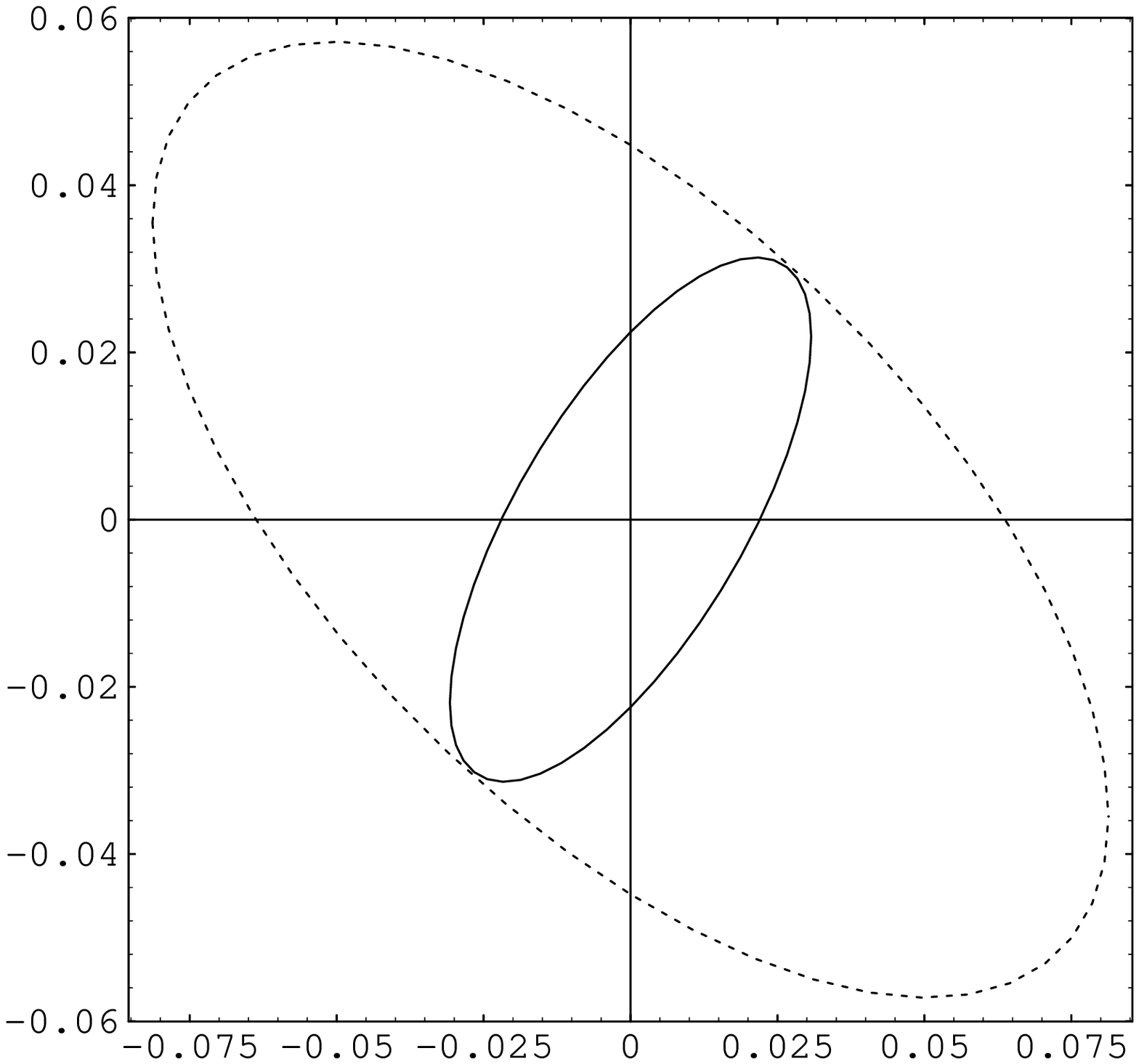,height=9cm}
\]
\vspace{-6.cm}\null\\
\hspace*{-1cm} \v \hspace{8cm} \v \\[3.2cm]
\hspace*{6cm} \x \hspace{8cm}  \x
\\
\hspace*{3.2cm} (c) \hspace{7.8cm}  (d)
\\[1.2cm]
\caption[1]{Observability limits in the 6-parameter case;
with polarized beams (a) (c), with unpolarized beams
(b) (d); from asymmetries alone (- - - -), from asymmetries
and the 3 integrated $\rho^{L+R}_{ij}$ matrix elements 
with a normalization uncertainty of
2\% (........), 20\% (----------).}
\end{figure}
\begin{figure}[p]
\vspace*{-2.cm}
\[
\epsfig{file=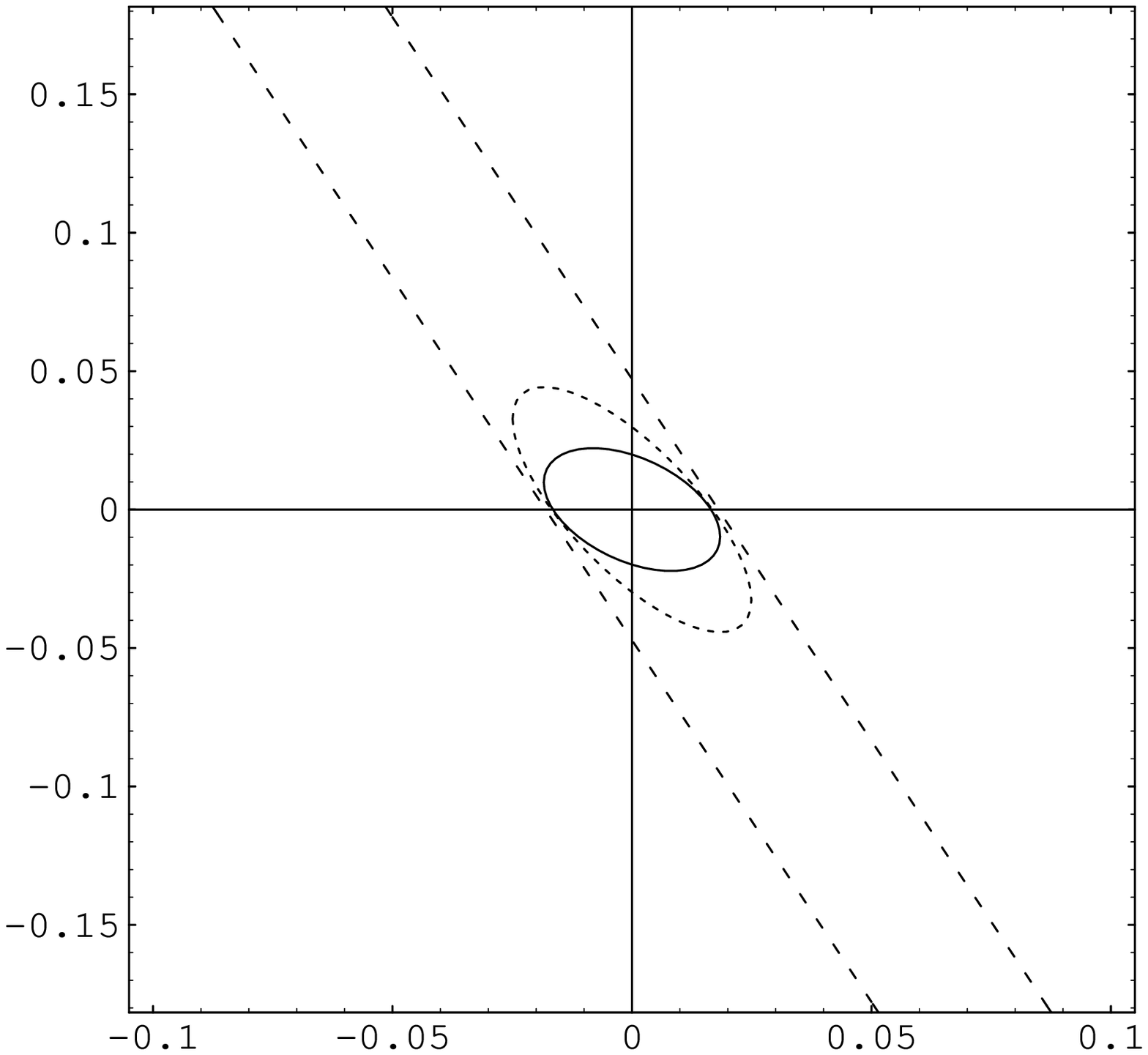,height=9cm}\hspace{2cm} 
\epsfig{file=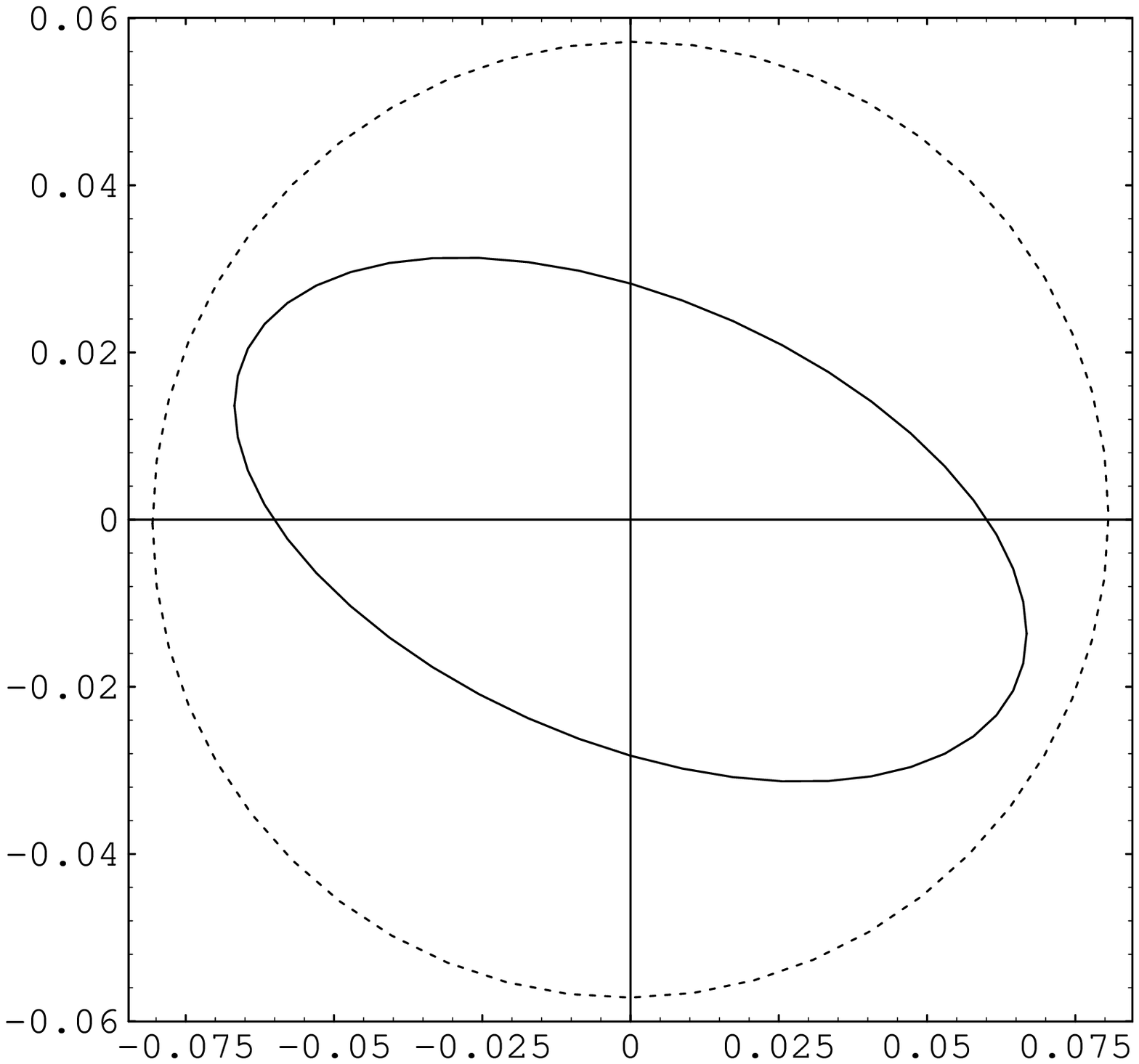,height=9cm}
\]
\vspace{-6.cm}\null\\
\hspace*{-1cm} \v \hspace{8cm} \v \\[3.2cm]
\hspace*{6cm} \u \hspace{8cm}  \u
\\
\hspace*{3.2cm} (e) \hspace{7.8cm}  (f)
\[
\epsfig{file=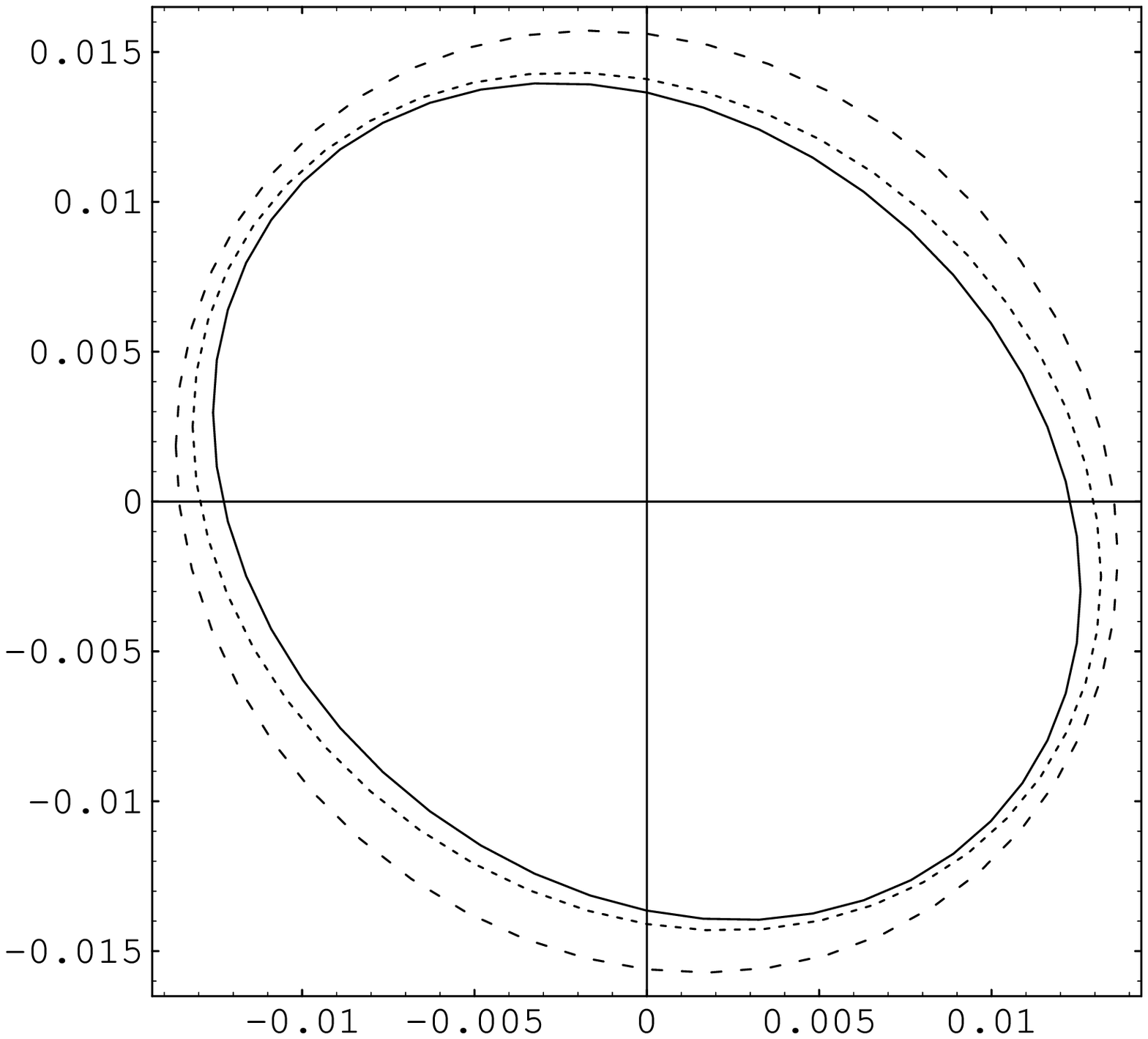,height=9cm}\hspace{2cm} 
\epsfig{file=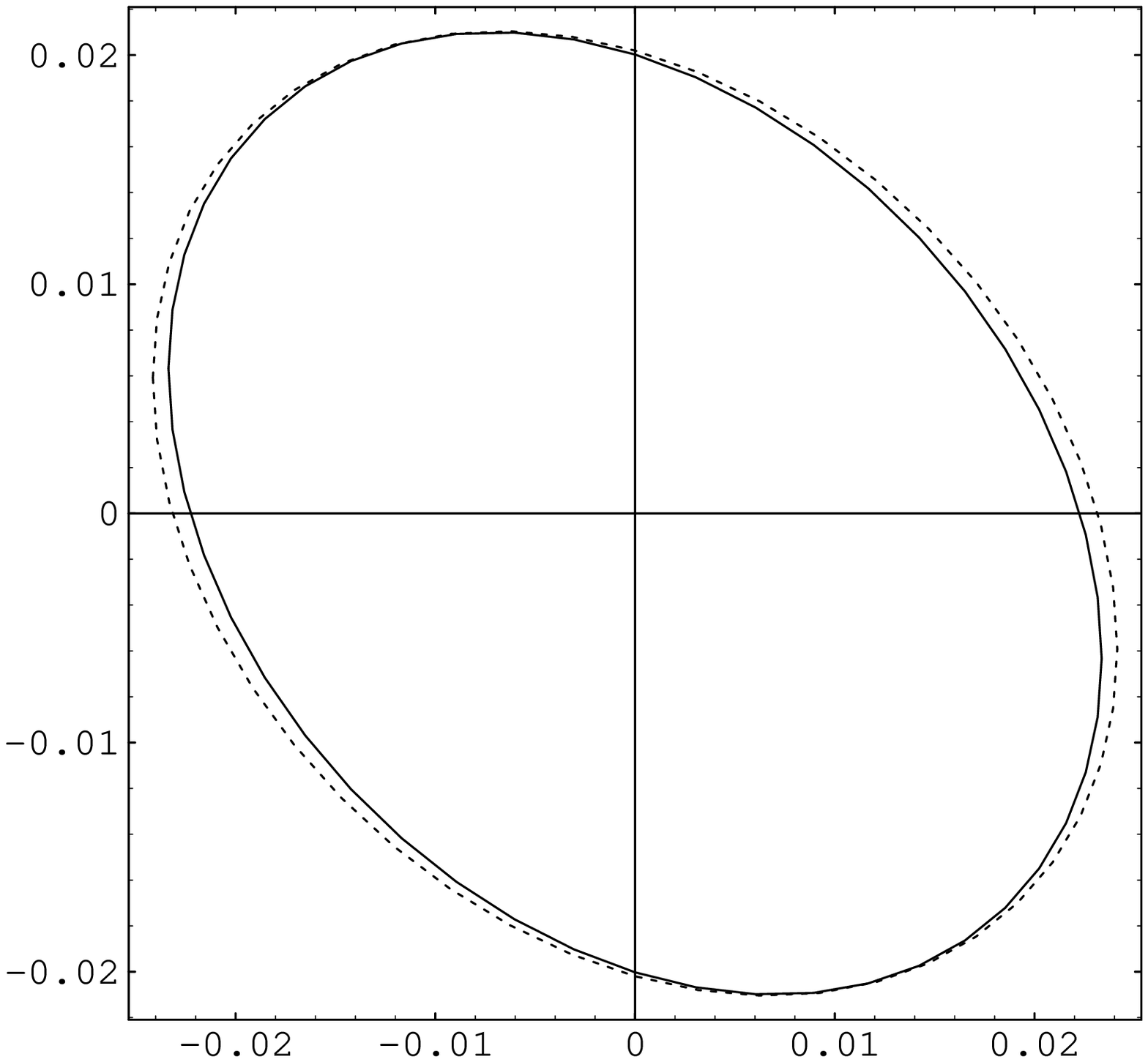,height=9cm}
\]
\vspace{-6.cm}\null\\
\hspace*{-1cm} \z \hspace{8cm} \z \\[3.2cm]
\hspace*{6cm} \y \hspace{8cm}  \y
\\
\hspace*{3.2cm} (g) \hspace{7.8cm}  (h)
\\[1.2cm]
\caption[1]{Observability limits in the 6-parameter case;
with polarized beams (e) (g), with unpolarized beams
(f) (h); from asymmetries alone (- - - -), from asymmetries
and the 3 integrated $\rho^{L+R}_{ij}$ matrix elements 
with a normalization uncertainty of
2\% (........), 20\% (----------).}
\end{figure}
\begin{figure}[p]
\vspace*{-2.cm}
\[
\epsfig{file=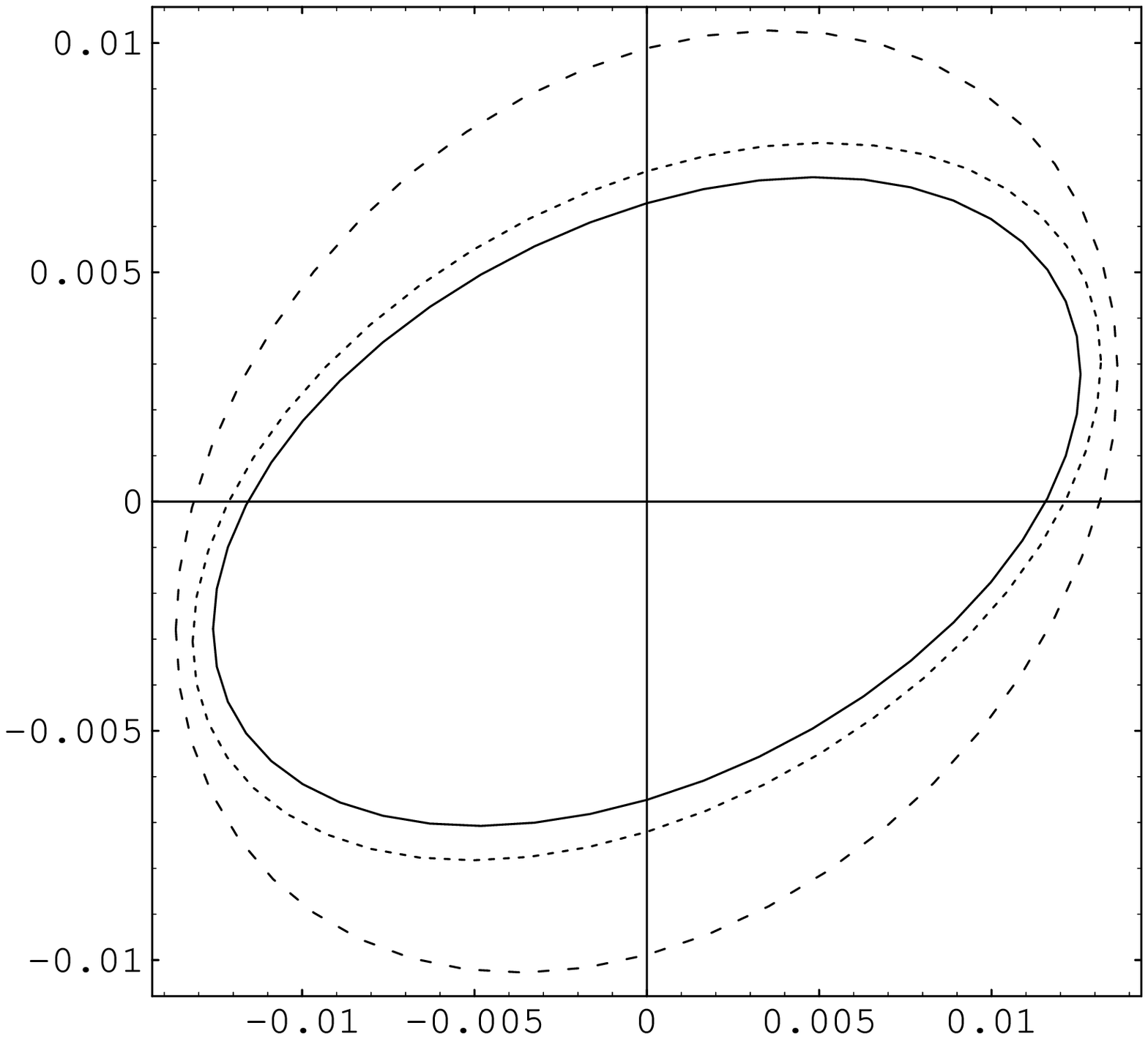,height=9cm}\hspace{2cm} 
\epsfig{file=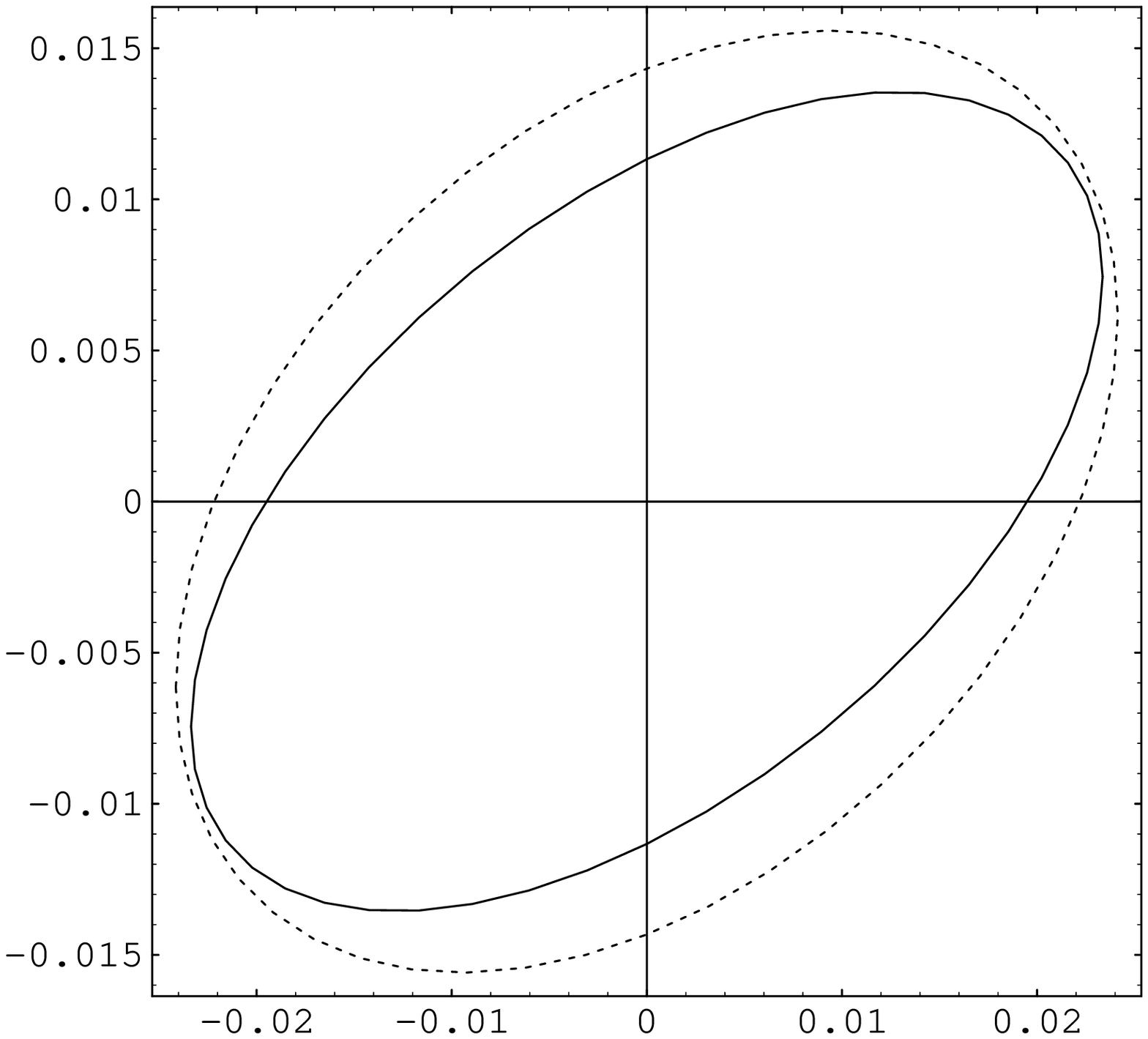,height=9cm}
\]
\vspace{-6.cm}\null\\
\hspace*{-1cm} \w \hspace{8cm} \w \\[3.2cm]
\hspace*{6cm} \y \hspace{8cm}  \y
\\
\hspace*{3.2cm} (i) \hspace{7.8cm}  (j)
\[
\epsfig{file=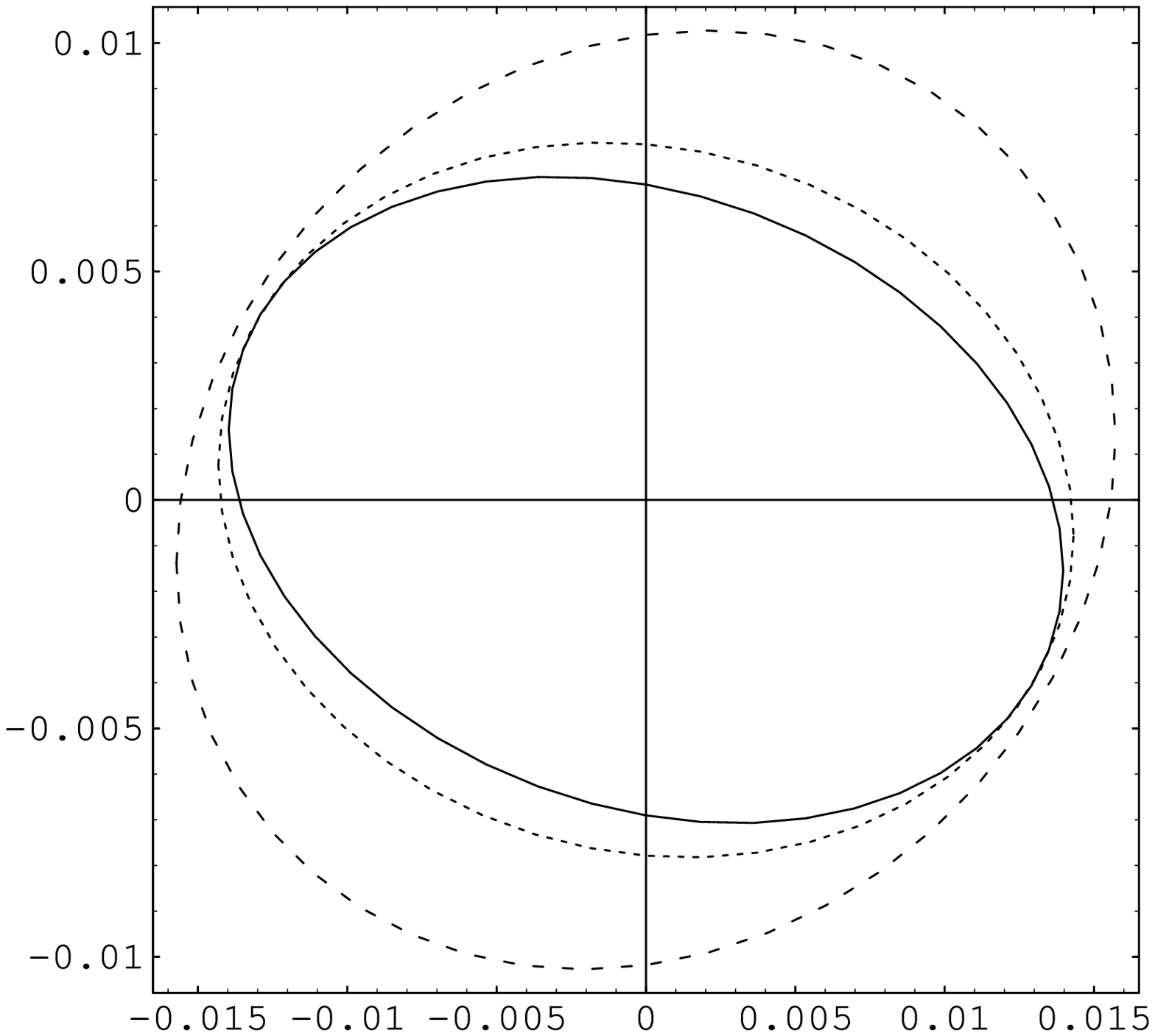,height=9cm}\hspace{2cm} 
\epsfig{file=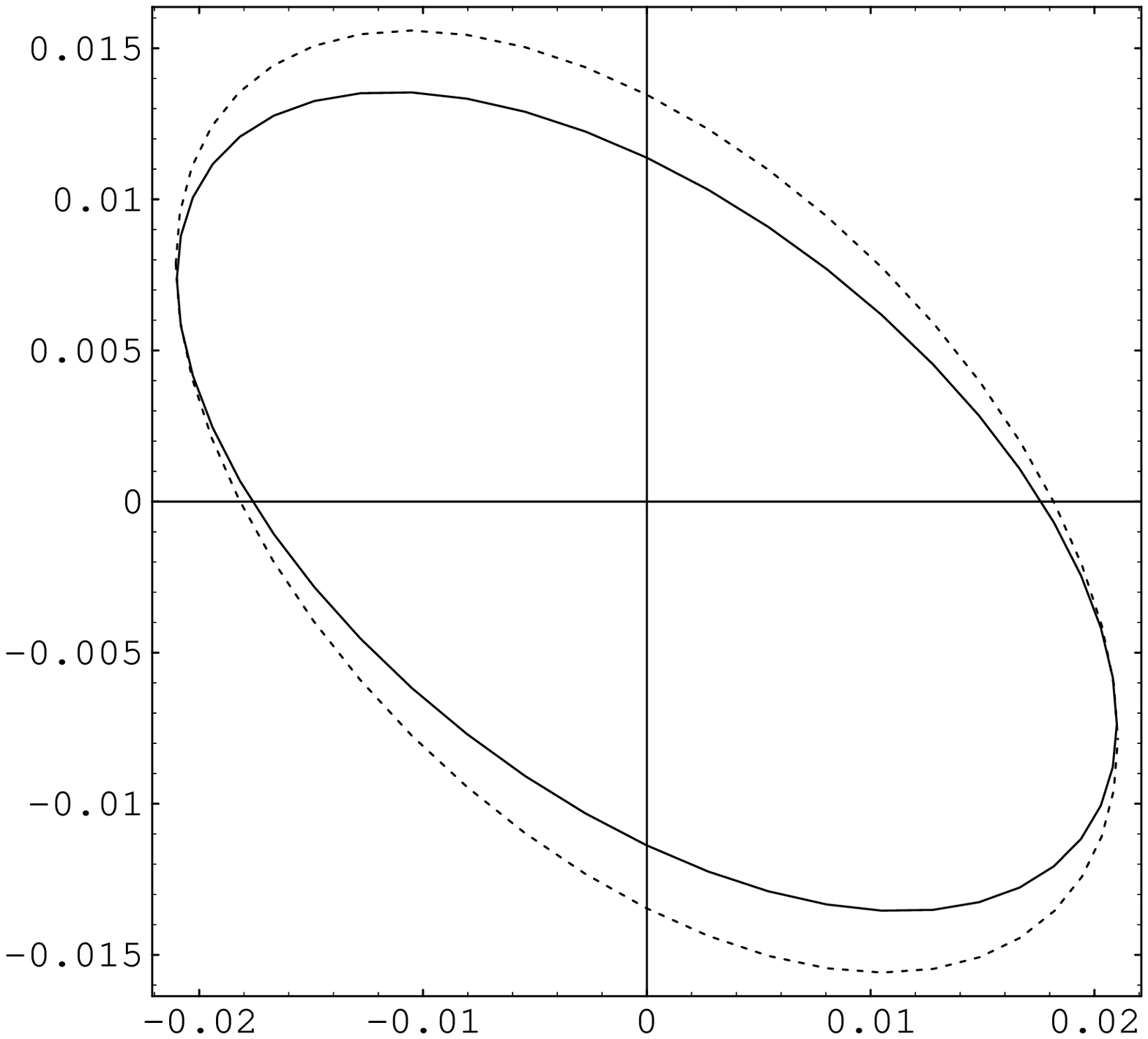,height=9cm}
\]
\vspace{-6.cm}\null\\
\hspace*{-1cm} \w \hspace{8cm} \w \\[3.2cm]
\hspace*{6cm} \z \hspace{8cm}  \z
\\
\hspace*{3.2cm} (k) \hspace{7.8cm}  (l)
\\[1.2cm]
\caption[1]{Observability limits in the 6-parameter case;
with polarized beams (i) (k), with unpolarized beams
(j) (l); from asymmetries alone (- - - -), from asymmetries
and the 3 integrated $\rho^{L+R}_{ij}$ matrix elements 
with a normalization uncertainty of
2\% (........), 20\% (----------).}
\end{figure}

\newpage

\renewcommand{\arraystretch}{1.2}
\begin{table}[p]
\begin{small}
\begin{tabular}{|c|c|c|c|c|} \hline
\multicolumn{5}{|c|}{Table 1: Sensitivity limits for 1 free parameter,
polarized/unpolarized cases}\\
\multicolumn{1}{|c|}{Operator} &
  \multicolumn{1}{|c|}{$\sqrt{s}=0.5$ TeV} &
   \multicolumn{1}{|c|}{$\sqrt{s}=1$ TeV} &
     \multicolumn{1}{|c|}{$\sqrt{s}=2$ TeV} &
       \multicolumn{1}{|c|}{other constraints}
          \\[0.1cm] \hline
  $\O_{qt}$ & 0.53/1.62& 0.26/0.53 &
0.18/0.31& $-0.14\pm0.07^{(b)}$\\
   & (0.98)/(0.56)& (1.41)/(0.99) &
(1.68)/(1.29)& \\[0.1cm]\hline
 
  $\O^{(8)}_{qt}$ & 0.10/0.30 & 0.049/0.099
&0.034/0.057 &$-0.027\pm0.013^{(b)}$
\\
  & (2.47)/(1.42) & (3.55)/(2.49)
&(4.24)/(3.26) & 
\\[0.1cm]\hline
  $\O_{tt}$ & 0.064/0.11 & 0.017/0.039 &
0.010/0.026 & ----- \\
  &(3.00)/(2.32) & (5.87)/(3.85) &
(7.46)/(4.70) & \\[0.1cm]\hline
  $\O_{tb}$ & 0.14/0.36 & 0.043/0.11 &
0.027/0.071 & $-0.13\pm0.06^{(b)}$\\
  & (2.33)/(1.46) & (4.24)/(2.64) &
(5.30)/(3.30) & \\[0.1cm]\hline
$\O_{t2}$ & 0.010/0.023& 0.0090/0.018 &
0.0089/0.017 &$0.01^{(a)}$; $0.14\pm0.07^{(b)}$ \\
 & (11.57)/(7.60)& (12.18)/(8.62) &
(12.24)/(8.75) &  \\[0.1cm]\hline
  $\O_{Dt}$ &0.039/0.093&
0.011/0.018&0.0052/0.0071&$0.03^{(a)}$; $-0.06\pm0.03^{(b)}$ \\
   &(2.84)/(1.85)&
(5.27)/(4.15)&(7.84)/(6.68)& \\[0.1cm]\hline
$\O_{tW\Phi}$ & 0.0010/0.0021 &
0.00067/0.0010& 0.00043/0.00056 &$0.014^{(a)}$
\\
 & (42.67)/(29.97) &
(53.14)/(42.73)& (66.16)/(58.10) &
\\[0.1cm]\hline
  $\O_{tB\Phi}$ & 0.0011/0.0027 &
0.00079/0.0015 &0.00060/0.0012 & $0.013^{(a)}$ \\
 & (41.63)/(26.52) &
(48.82)/(36.11) &(55.89)/(39.80) & \\
[0.1cm]\hline
$\O_{tG\Phi}$ & 0.027/0.029 & 0.023/0.025
 & 0.045/0.047 & ----- \\
 & (7.86)/(7.30) & (9.08)/(8.54)
 & (4.71)/(4.52) & \\[0.1cm]\hline
$\O_W$ & 0.065/0.13 & 0.021/0.045 &
0.014/0.030 & $0.1^{(c)}$ \\
 & (1.37)/(0.95) & (2.38)/(1.65) &
(2.95)/(2.02) & \\[0.1cm]\hline
$\O_{W\Phi}$ & 0.11/0.22 & 0.036/0.075 &
0.023/0.050 & $0.1^{(c)}$ \\
& (1.35)/(0.94) & (2.35)/(1.63) &
(2.91)/(1.99) &  \\[0.1cm]\hline
$\O_{B\Phi}$ & 0.071/0.14 & 0.020/0.043 &
0.012/0.028 & $0.1^{(c)}$ \\
& ((2.98)/(2.09) & (5.66)/(3.81) &
(7.14)/(4.76) &  \\[0.1cm]\hline
$\O_{WW}$ & 0.28/0.56 & 0.29/0.45 &
0.51/0.66 & $0.015^{(c)}$ \\
& (1.56)/(1.10) & (1.52)/(1.22) &
(1.15)/(1.01) &  \\[0.1cm]\hline
$\O_{BB}$ & 0.32/0.78 & 0.37/0.69
&0.77/1.53  & $0.05^{(c)}$ \\
 & (1.99)/(1.27) & (1.83)/(1.35)
&(1.27)/(0.91)  &  \\[0.1cm]\hline
$\O_{\Phi 2}$ & 0.57/0.68 &  0.68/0.81 & 1.74/2.08& $0.01^{(c)}$  \\
\hline
\end{tabular}
\end{small}
\end{table}


\end{document}